\documentclass[%
 aip,
 amsmath,amssymb,
 reprint,%
]{revtex4-1}

\usepackage{graphicx}
\usepackage{dcolumn}
\usepackage{bm}
\usepackage{color}

\usepackage[utf8]{inputenc}
\usepackage[T1]{fontenc}
\usepackage{mathptmx}
\usepackage{etoolbox}

\makeatletter
\def\@email#1#2{%
 \endgroup
 \patchcmd{\titleblock@produce}
  {\frontmatter@RRAPformat}
  {\frontmatter@RRAPformat{\produce@RRAP{*#1\href{mailto:#2}{#2}}}\frontmatter@RRAPformat}
  {}{}
}%
\makeatother
\begin{document}
\preprint{AIP/123-QED}
\title[Plasma wakefields in microwave waveguides]{Particle-in-cell simulations of plasma wakefield formation in microwave waveguides}
\author{Jesús E. López}
\affiliation{School of Physics, Universidad Industrial de Santander, Bucaramanga, Colombia, 680002}
\author{Eduardo A. Orozco}
\email{eaorozco@uis.edu.co}
\affiliation{School of Physics, Universidad Industrial de Santander, Bucaramanga, Colombia, 680002}
\date{\today}
\begin{abstract}
The acceleration of charged particles is fundamental not only for experimental studies in particle physics but also for applications in fields such as semiconductor manufacturing and medical therapies. However, conventional accelerators face limitations due to their large size, driven by low acceleration gradients. Plasma-based accelerators have emerged as a promising alternative, offering ultrahigh acceleration gradients, though their implementation is often limited by the need for high-intensity, femtosecond laser systems and sophisticated diagnostics. As a more accessible alternative, the use of microwave pulses to excite plasma wakefields in waveguides filled with low-density plasma has gained attention. In this study, we perform three-dimensional particle-in-cell simulations to investigate the formation and structure of electrostatic wakefields driven by short microwave pulses in rectangular plasma waveguides. The results establish a theoretical basis for evaluating the feasibility and potential applications of microwave-driven plasma acceleration schemes.
\end{abstract}
\maketitle
\section{Introduction}
\noindent Plasma-based acceleration has emerged as a significant area of research in plasma physics, attracting sustained interest from both experimental and theoretical efforts, including numerical simulations \cite{RevModPhys.81.1229,malka2012laser}. Unlike conventional radiofrequency (RF) accelerators, limited by electrical breakdown thresholds and typically offering acceleration gradients below 100~MV/m \cite{wangler2008rf,litos2014high}, plasma-based accelerators can sustain electric fields on the order of GV/m \cite{tajima1979laser,RevModPhys.81.1229}. This capability enables the acceleration of charged particles to relativistic energies over much shorter distances, opening possibilities for compact accelerator designs. This mechanism commonly relies on the interaction of an intense, short-duration laser pulse with an underdense plasma, which excites a longitudinal electrostatic wave, known as a Langmuir wave or plasma wakefield. The wakefield acts as a co-moving accelerating structure that trails behind the driving pulse. Experimental implementations of laser wakefield acceleration (LWFA) typically employ near-petawatt (PW) laser pulses with durations of tens of femtoseconds interacting with plasmas of densities around $10^{17}$~cm$^{-3}$. Despite the remarkable acceleration gradients achieved, the high cost and complexity of ultrafast, high-intensity laser systems remain substantial barriers to widespread implementation~\cite{albert2016applications,cakir2019brief}.

\vspace{1pc}

\noindent As a potentially more accessible alternative, microwave-driven plasma wakefield acceleration has recently garnered attention. Arya and Malik (2008) developed a one-dimensional, quasi-nonlinear analytical model predicting wakefield excitation in rectangular waveguides filled with low-density plasma~\cite{aria2008wakefield}. According to their analytical model, a microwave pulse with a frequency of 8~GHz, a temporal duration of 0.8~ns, and an intensity of $1 \times 10^{9}$~W/m$^2$ is capable of inducing wakefield excitation in a rectangular waveguide containing plasma with an electron density of $n = 1.8 \times 10^{16}$~m$^{-3}$.
 Crucially, their model emphasized the need to operate above the waveguide’s cut-off frequency, determined by its transverse dimensions and the plasma dielectric response, to ensure efficient propagation and avoid deleterious instabilities. Recent experimental progress has further strengthened interest in this approach. Notably, Cao and collaborators have developed high-power microwave sources designed for wakefield excitation in cylindrical, plasma-filled waveguides~\cite{bliokh2017wakefield,krasik2019experiments,cao2019wakefield}. Their goal is to produce compact acceleration stages for applications such as X-ray generation. Most recently, experimental evidence has confirmed the generation of plasma waves with electric field amplitudes up to 20~kV/cm~\cite{cao2024direct}, demonstrating the viability of microwave-driven wakefield generation.

\vspace{1pc}

\noindent Cao and collaborators has combined experimental development and computational modeling to explore microwave-driven plasma wakefield excitation. They experimentally developed high-power microwave (HPM) sources based on cylindrical wire-array waveguides, capable of generating pulses with peak powers near 1.2~GW, durations around 0.4~ns, and central frequencies close to 10~GHz. Using these parameters, they employed particle-in-cell (PIC) simulations to show that such pulses can successfully excite wakefields via the cylindrical TM$_{01}$ mode in plasma-filled waveguides. These results highlight the potential of microwave technology for compact plasma-based accelerators. In contrast, the present study focuses on rectangular waveguide configurations and TE$_{10}$-mode pulses, which are standard in conventional microwave systems. By systematically varying key parameters, including pulse duration, frequency, power, waveguide geometry, and plasma density, we investigate the wakefield response and analyze how each factor influences the amplitude and structure of the resulting plasma wave. To achieve this, we employ fully electromagnetic, three-dimensional PIC simulations to resolve the nonlinear and kinetic dynamics of the plasma–microwave pulse interaction.

\vspace{1pc}

\noindent The findings contribute to the theoretical groundwork necessary to evaluate the viability of this acceleration scheme and guide future experimental implementations in compact, cost-effective accelerator systems.
\section{Theoretical Formalism}
\noindent Plasma wakefields are collective oscillations that arise when an external driver, such as an electromagnetic pulse, propagates through a plasma medium \cite{esarey2009physics,malka2012laser}. As the driver advances, it displaces electrons from their equilibrium positions, creating a local displacement in charge density. This space-charge separation generates Langmuir waves, longitudinal oscillations of the electron component, that trail behind the driver with a phase velocity approximately equal to its group velocity \cite{jackson2021classical,tajima1979laser}. The resulting wake structure supports strong electric fields, with amplitudes typically on the order of
\begin{equation}
    E_{\text{wake}} \sim \frac{m_e c \omega_p}{e},
    \label{E_wake}
\end{equation}
where the plasma frequency is given by
\begin{equation}
    \omega_p^2 = \frac{n_e e^2}{\varepsilon_0 m_e},
    \label{EQ_plasma_frequency}
\end{equation}
with \(n_e\) the electron density, \(e\) the elementary charge, \(\varepsilon_0\) the vacuum permittivity, and \(m_e\) the electron mass. The associated plasma wavelength, \(\lambda_p = 2\pi v_g / \omega_p\), defines the spatial period of the wake, where \(v_g\) is the group velocity of the driving pulse.

\vspace{1pc}

\noindent
The amplitude and structure of the wakefield depend critically on the properties of the driver: its intensity, duration, and frequency, as well as the background plasma density. In the present study, wakefields are excited by the fundamental TE\(_{10}\) mode of a microwave pulse propagating through a rectangular waveguide filled with low-density plasma.
\subsection{Governing Equations}
\begin{figure*}[ht]
    \centering
    \includegraphics[width=0.90\linewidth]{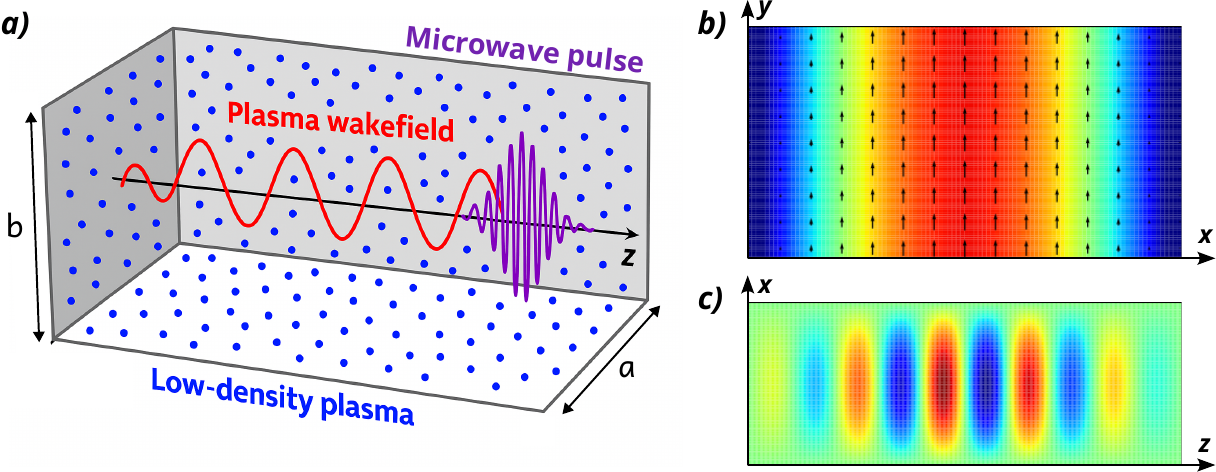}
    \caption{\label{FIG_scheme}(a) Schematic representation of the rectangular waveguide filled with plasma. The TE$_{10}$ microwave pulse propagates along the $z$-direction. Electric field distribution of the TE$_{10}$ mode in the transverse planes: (b) $z \equiv \text{cte}$ and (c) $y \equiv \text{cte}$.
    }
\end{figure*}

\noindent The primary objective of this work is to characterize the wakefield generated by a TE\(_{10}\) microwave pulse in a rectangular plasma-filled waveguide. A Schematic representation of the rectangular waveguide filled with plasma is shown in FIG~\ref{FIG_scheme}(a), with the pulse propagating along the \(z\)-axis. The corresponding electric field distribution, shown in FIG~\ref{FIG_scheme}(b), displays a half-wavelength sinusoidal profile along the $x$-direction, with no variation along $y$-axis, consistent with the TE$_{10}$ mode supported by the rectangular waveguide and the imposed metallic boundary conditions \cite{pozar2012microwave,jackson2021classical}. Plasma behavior is governed by the interaction between electromagnetic field and the plasma response, described by the charge density \(\rho\) and current density \(\mathbf{J}\). The problem is inherently nonlinear and self-consistent: the electromagnetic pulse modifies the plasma state, and in turn, the plasma affects the evolution of the fields. To model this interaction accurately in the low-density kinetic regime of interest, we adopt a first-principles approach based on the Vlasov–Maxwell system of equations \cite{pukhov2015particle}.

\vspace{1pc}

\noindent The Vlasov equation describes the evolution of the distribution function \(f_\alpha(\mathbf{r}, \mathbf{p}, t)\) for each species \(\alpha\),
\begin{equation}
    \frac{\partial f_{\alpha}}{\partial t} 
    + \frac{\mathbf{p}}{\gamma m_{\alpha}} \cdot \nabla_{\mathbf{r}} f_{\alpha} 
    + q_{\alpha} \left( \mathbf{E} + \frac{\mathbf{p}}{\gamma m_{\alpha}} \times \mathbf{B} \right) \cdot \nabla_{\mathbf{p}} f_{\alpha} = 0,
    \label{EQ_vlasov}
\end{equation}
with \(q_\alpha\) and \(m_\alpha\) the charge and mass of the species, and \(\gamma\) the Lorentz factor. The fields, $\mathbf{E}$ and $\mathbf{B}$, obey Maxwell’s equations:
\begin{align}
    \nabla\cdot\mathbf{E} &= \frac{\rho}{\varepsilon_0}, \label{EQ_gauss_E} \\
    \nabla\cdot\mathbf{B} &= 0, \label{EQ_gauss_B} \\
    \nabla\times\mathbf{E} &= -\frac{\partial \mathbf{B}}{\partial t}, \label{EQ_faraday} \\
    \nabla\times\mathbf{B} &= \mu_0\mathbf{J} + \mu_0\varepsilon_0\frac{\partial \mathbf{E}}{\partial t}, \label{EQ_ampere-maxwell}
\end{align}
with current and charge densities:
\begin{align}
    \rho(\mathbf{r}, t) &= \sum_\alpha q_\alpha \int f_\alpha(\mathbf{r}, \mathbf{p}, t) \, d\mathbf{p}, \label{EQ_charge_density} \\
    \mathbf{J}(\mathbf{r}, t) &= \sum_\alpha q_\alpha \int \frac{\mathbf{p}}{\gamma m_\alpha} f_\alpha(\mathbf{r}, \mathbf{p}, t) \, d\mathbf{p}. \label{EQ_current_density}
\end{align}
\subsection{Numerical scheme}
\noindent To solve the Vlasov–Maxwell system numerically, we employ fully electromagnetic, three-dimensional Particle-in-Cell (PIC) simulations. The PIC method is ideally suited for studying plasma dynamics in kinetic and nonlinear regimes~\cite{birdsall2004plasma}, as it solves the coupled evolution of particles and electromagnetic fields in a self-consistent manner. In this framework, the plasma is represented numerically by discrete elements of phase space known as macroparticles. Each macroparticle represents a large number of real particles that are close together in phase space. The dynamics of each macroparticle is governed by the relativistic Newton–Lorentz equations,~\cite{hockney1988computer,birdsall2004plasma,lapenta2015kinetic}
\begin{align}
    \frac{d\mathbf{p}_i}{dt} &= q_i \left[ \mathbf{E}(\mathbf{r}_i, t) + \mathbf{v}_i \times \mathbf{B}(\mathbf{r}_i, t) \right], \label{EQ_particle_motion} \\
    \frac{d\mathbf{r}_i}{dt} &= \frac{\mathbf{p}_i}{\gamma m_i}, \label{EQ_particle_position}
\end{align}
where \(\mathbf{p}_i = \gamma m_i \mathbf{v}_i\) is the relativistic momentum, and \(\gamma\) is the Lorentz factor.

\vspace{1pc}

\noindent The equations of motion for the macroparticles are integrated using the Boris algorithm~\cite{boris1970relativistic}, a second-order leapfrog scheme that provides accurate conservation of energy and momentum, even in the presence of strong electromagnetic fields. The field equations are solved using the finite-difference time-domain (FDTD) method on a staggered Yee grid~\cite{yee1966numerical}, ensuring second-order accuracy and preserving the divergence-free condition \(\nabla \cdot \mathbf{B} = 0\). If charge conservation is enforced numerically,
\begin{equation}
    \nabla \cdot \mathbf{J} + \frac{\partial \rho}{\partial t} = 0,
    \label{EQ_charge_continuity}
\end{equation}
then Gauss's law for \(\mathbf{E}\) is preserved throughout the simulation without explicit correction.

\vspace{1pc}

\noindent We employ the Umeda scheme for charge and current deposition, ensuring exact charge conservation~\cite{umeda2003new}. The simulation domain consists of a rectangular waveguide with metallic (PEC) walls at \(x = 0\), \(x = a\), \(y = 0\), and \(y = b\), filled with homogeneous, low-density plasma. A TE\(_{10}\) microwave pulse is initialized at \(z = 0\), and its propagation is monitored in time. The output boundary in \(z\) includes absorbing conditions to prevent unphysical reflections~\cite{berenger1994perfectly}.

\vspace{1pc}

\noindent For the macroparticles, a hybrid boundary condition is implemented to suppress artificial particle reflections and minimize numerical noise near metallic surfaces. A buffer region comprising three grid cells is defined adjacent to the transverse boundaries. Within this region, particles experience a weak, spatially dependent damping force proportional to their velocity. Particles with velocity magnitudes below a threshold (\(v < 0.01\,c\)) are removed from the simulation domain, while high-energy particles are smoothly decelerated and partially reflected. This treatment helps prevent unphysical particle accumulation at the walls and contributes to the overall stability of the simulation.

\vspace{1pc}

\noindent This computational framework captures the nonlinear and kinetic processes governing plasma wakefield formation in microwave-driven waveguides, enabling detailed exploration of wake structures under various physical parameters.
\section{Results and Discussion}
\noindent The results of the three-dimensional PIC simulations are presented and discussed, with emphasis on the influence of key parameters, such as pulse duration, intensity, and waveguide geometry, on the excitation and structure of the resulting plasma wakefield. The selection of simulation parameters is guided by experimental studies conducted by Cao and collaborators~\cite{bliokh2017wakefield,krasik2019experiments,cao2019wakefield,cao2024direct}, which employed low-density plasmas with electron densities on the order of \(n_e \sim 10^{16}~\text{m}^{-3}\). In those works, the driving microwave pulses featured Gaussian envelopes, central frequencies in the gigahertz range, and full-width at half-maximum (FWHM) durations of approximately 0.5~ns. The present simulations adopt these parameters to maintain consistency with experimentally realizable conditions and to enable meaningful comparisons between theory and observation. By systematically varying the relevant physical parameters, this study examines how the wakefield amplitude and spatial coherence respond to changes in the system configuration. Particular attention is devoted to the role of transverse confinement imposed by the waveguide geometry and its interplay with the plasma's dielectric response, which critically influences the efficiency and structure of the excited wakefield. These insights contribute to the design of compact, microwave-driven plasma-based accelerators.

\vspace{1pc}

\begin{figure*}[ht]
    \centering
    \includegraphics[width=0.95\linewidth]{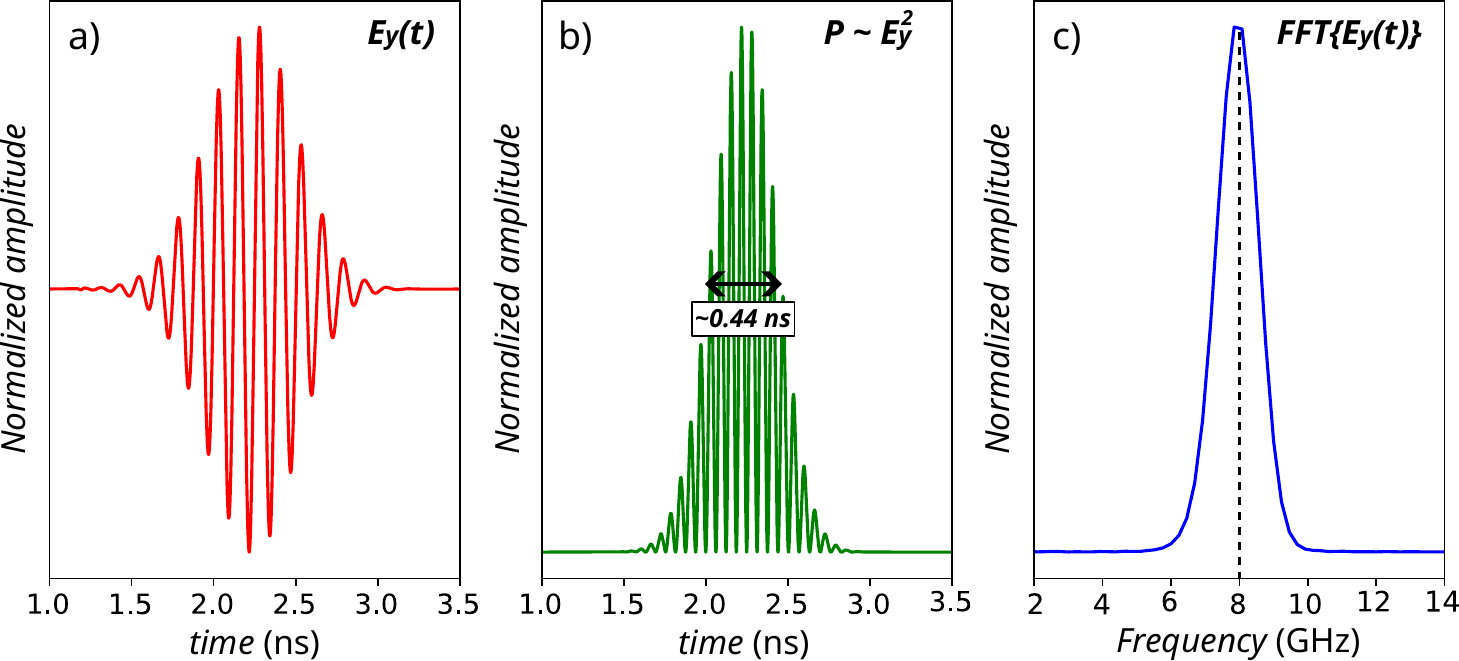}
    \caption{\label{Fig_MW_pulse}Temporal and spectral characterization of the injected microwave pulse. The Fig.(a) shows the electric field component \(E_y(t)\) at the point \( (\frac{a}{2},\frac{b}{2},\frac{Lz}{3}) \). The sinusoidal oscillations modulated by a Gaussian envelope confirm the expected temporal structure. The Fig.(b) shows the instantaneous power profile, proportional to \(E_y^2(t)\), from which the pulse duration is calculated. The Fig.(c) displays the frequency spectrum, FFT\(\{E_y(t)\}\), confirming a Gaussian-like profile centered at 8~GHz. These results validate the correct initialization of the microwave pulse in both time and frequency domains.}
\end{figure*}

\noindent As a reference case, we consider a homogeneous plasma with density \(n_e = 1.8\times10^{16}~\text{m}^{-3}\) was confined within a rectangular waveguide of dimensions \(a = 3~\text{cm}\) and \(b = 2.4~\text{cm}\), corresponding to an aspect ratio \(b/a = 0.8\). The plasma was driven by a microwave pulse with central frequency 8.0~GHz, peak power of 0.25~GW, and a FWHM duration of approximately 0.44~ns. FIG~\ref{Fig_MW_pulse} illustrates the correct initialization of the TE$_{10}$ mode pulse, showing good agreement with the expected spectral and temporal characteristics. The pulse maintains its Gaussian envelope and desired carrier frequency as it propagates through the plasma-filled waveguide.

\vspace{1pc}

\begin{figure*}[ht]
    \centering
    \includegraphics[width=1.0\linewidth]{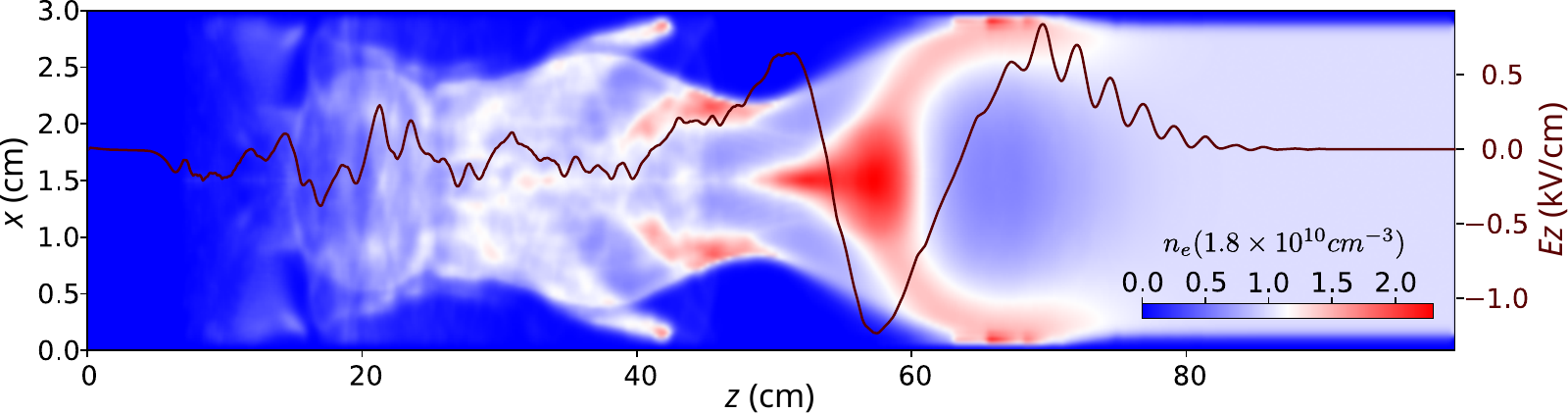}
    \caption{\label{Fig_Ez_sim_08}Spatial structure of the plasma wakefield generated behind the microwave pulse. The colormap shows the perturbed electron density \(n_e\) in the midplane \(y = b/2\), while the brown curve shows the longitudinal electric field \(E_z(z)\) along the axis (\(x = a/2\), \(y = b/2\)). The presence of a non-zero \(E_z\), despite the transverse nature of the TE$_{10}$ mode, confirms the excitation of a Langmuir-type electrostatic mode. The initial negative peak in \(E_z\), exceeding 1~kV/m, indicates a region, in \( 54~cm \lesssim z \lesssim 63~cm\), suitable for electron acceleration.}
\end{figure*}

\noindent The simulation results demonstrate the successful excitation of an electrostatic plasma wave, a wakefield, trailing the injected microwave pulse. FIG~\ref{Fig_Ez_sim_08} shows a colormap of the perturbed electron density \(n_e\) in the transverse midplane (\(y = b/2\)), along with the longitudinal electric field \(E_z(z)\) along the axis (\(x = a/2\), \(y = b/2\)). The longitudinal oscillations in both charge density and electric field confirm the formation of a plasma wake.

\vspace{1pc}

\noindent
It is worth emphasizing that the TE$_{10}$ mode in vacuum is characterized by a purely transverse electric field and zero longitudinal component. Therefore, the presence of a non-zero \(E_z\) in the plasma is a manifestation of wakefield excitation, specifically a Langmuir wave driven by the ponderomotive force associated with the transverse pulse envelope. This makes TE modes particularly suitable for wakefield studies, as any observed longitudinal electric field can be unambiguously attributed to the plasma response. The wakefield exhibits a non-uniform amplitude along the propagation axis. The first peak reaches an amplitude slightly above \(1~\text{kV/m}\), followed by damped oscillations with peak spacings of approximately \(\lambda_p/2\), where \(\lambda_p\) is the local plasma wavelength. Notably, the first peak is negative, corresponding to an accelerating field for electrons in the direction of pulse propagation.

\vspace{1pc}

\noindent To investigate the influence of waveguide geometry on wakefield generation, two sets of simulations were performed. In the first set, the waveguide width \(a\) was systematically varied between 3.0 and 7.0\,cm, while maintaining a constant aspect ratio of \(b/a = 0.8\). Under this constraint, both the waveguide height \(b\) and the transverse cross-sectional area \(A_T = ab\) increased proportionally with \(a\). The plasma density was fixed at \(n_e = 1.8 \times 10^{16}~\text{m}^{-3}\), while the microwave pulse parameters: central frequency \(f_0 = 8\,\text{GHz}\), peak power of 0.25\,GW, and pulse duration of 0.44\,ns, were kept unchanged throughout the simulations. The corresponding peak wakefield amplitudes are summarized in TABLE~\ref{tab:AT_Ew}. The results exhibit a clear inverse correlation between the waveguide cross-sectional area and the wakefield amplitude. A power-law fit to the data yields the relation \(E_z = 24.05 \cdot A_T^{-1.5} + 0.03\), with a high correlation coefficient: \(R^2 = 0.9978\). This scaling show that increasing the transverse area reduces the electromagnetic energy density delivered by the pulse, leading to weaker plasma excitation and lower wakefield amplitudes.

\begin{table}
\caption{\label{tab:AT_Ew}Simulated wakefield amplitude \(E_z\) as a function of waveguide width \(a\) and corresponding cross-sectional area \(A_T = ab\). The plasma density, pulse frequency, duration, and power are held constant across all cases.}
\begin{ruledtabular}
\begin{tabular}{ccc}
\(a\) (cm) & \(A_T\) (cm\(^2\)) & \(E_z\) (kV/cm)\\
\hline
3.00 & 7.20  & 1.29 \\
4.00 & 12.80 & 0.52 \\
5.00 & 20.00 & 0.30 \\
6.00 & 28.80 & 0.19 \\
7.00 & 39.20 & 0.15 \\
\end{tabular}
\end{ruledtabular}
\end{table}

\vspace{1pc}

\noindent In the second set of simulations, the aspect ratio \(b/a\) was systematically varied between 0.5 and 0.9, while keeping the waveguide width fixed at \(a = 3.0\,\text{cm}\). This approach ensured that any variation in the cross-sectional area \(A_T\) originated exclusively from changes in the vertical dimension \(b\), thereby isolating its influence on wakefield formation. The results, summarized in TABLE~\ref{tab:ba_Ew}, indicate that the wakefield amplitude \(E_z\) remains nearly constant at approximately 1.29\,kV/cm across most configurations. A notable deviation is observed at \(b/a = 0.70\), where the amplitude increases to a peak value of 1.55\,kV/cm.

\begin{table}
\caption{\label{tab:ba_Ew}Wakefield amplitude \(E_z\) as a function of aspect ratio \(b/a\) with fixed waveguide width \(a = 3.0\) cm.}
\begin{ruledtabular}
\begin{tabular}{ccc}
\(b/a\) & \(A_T\) (cm\(^2\)) & \(E_w\) (kV/cm)\\
\hline
0.50 & 4.50 & 1.29 \\
0.60 & 5.40 & 1.29 \\
0.70 & 6.30 & 1.55 \\
0.80 & 7.20 & 1.29 \\
0.90 & 8.10 & 1.29 \\
\end{tabular}
\end{ruledtabular}
\end{table}

\vspace{1pc}

\begin{figure*}[ht]
    \centering
    \includegraphics[width=1.0\linewidth]{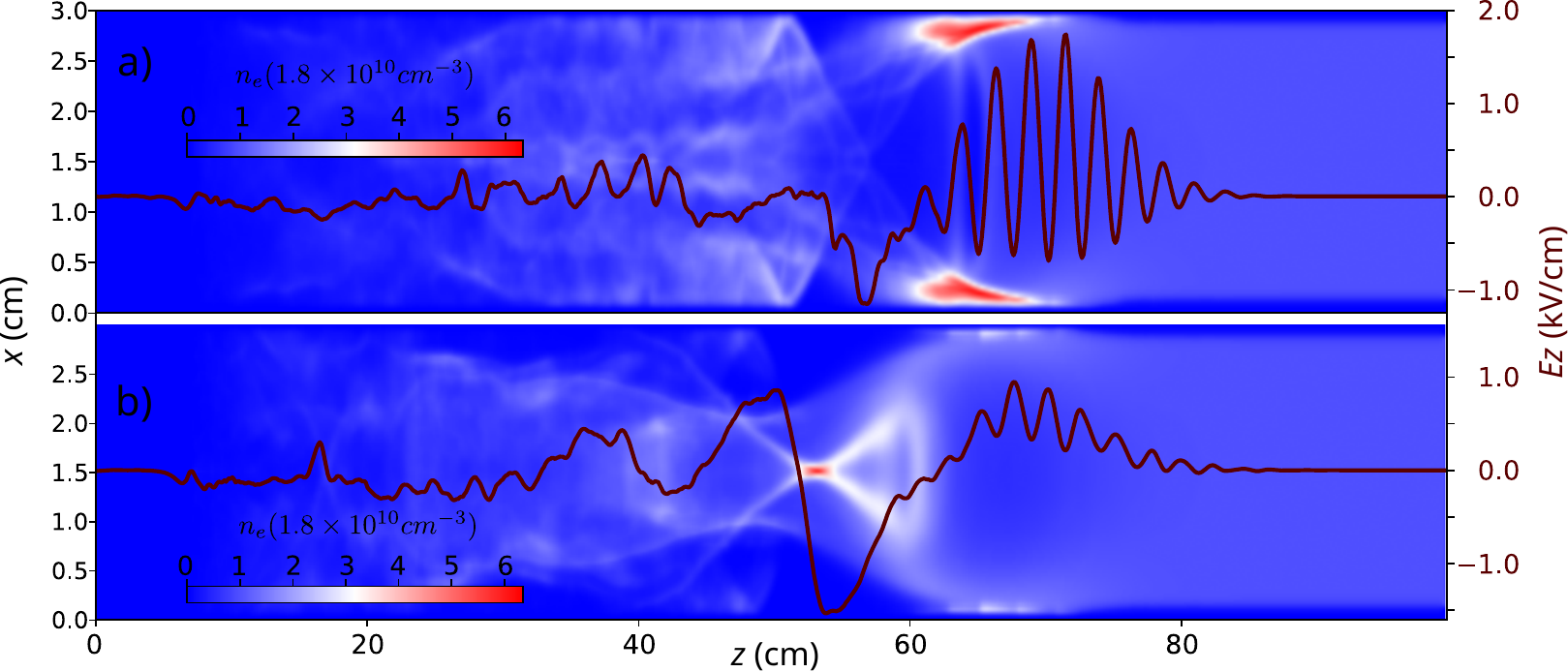}
    \caption{\label{Fig_Ez_vs_z_b} 
    Spatial structure of the plasma wakefield for two different waveguide aspect ratios: (a) \(b/a = 0.5\), and (b) \(b/a = 0.7\). Each panel shows the longitudinal electric field \(E_z(z)\) along the central axis (\(x = a/2\), \(y = b/2\)) and a colormap of the perturbed electron density \(n_e(x,z)\) in the central transverse plane \(y = b/2\). For the optimal configuration \(b/a = 0.7\), the electron density peaks in the central region behind the pulse, supporting strong wakefield excitation. In contrast, for \(b/a = 0.5\), the electron density is concentrated near the lateral walls of the waveguide, leading to a depleted central region and a substantially weaker wakefield. These results highlight the impact of waveguide geometry on both field excitation and plasma response.}
\end{figure*}

\noindent FIG~\ref{Fig_Ez_vs_z_b} presents the longitudinal electric field profiles \(E_z(z)\) obtained from two representative simulations listed in TABLE~\ref{tab:ba_Ew}, corresponding to aspect ratios \(b/a = 0.5\) and \(b/a = 0.7\). In addition to the \(E_z(z)\) profiles, the figure includes two-dimensional colormaps of the perturbed electron density \(n_e(x,z)\) in the central transverse plane (\(y = b/2\)). These results highlight the sensitivity of wakefield formation to the transverse geometry of the waveguide. Specifically, for the case \(b/a = 0.7\), which exhibits the largest wakefield amplitude, a pronounced electron density peak forms in the central region immediately behind the pulse, indicative of efficient charge displacement and coherent Langmuir wave excitation. In contrast, for \(b/a = 0.5\), electron accumulation is predominantly localized near the lateral walls, reducing the plasma response in the waveguide axis and thereby diminishing the strength and spatial coherence of the wakefield. In contrast, for the smaller aspect ratio \(b/a = 0.5\), the electron density is concentrated near the lateral walls, leaving the central axis partially depleted. This depletion correlates with a significantly weaker and less structured wakefield. These findings highlight the role of transverse plasma distribution in sustaining strong wakefield formation. These observations indicate that the waveguide width \(a\) plays a more significant role in determining the wakefield amplitude than the height \(b\). This conclusion is consistent with the field profile of the TE$_{10}$ mode, which exhibits a sinusoidal variation in the \(x\)-direction, defined by \(a\), and remains nearly uniform in the \(y\)-direction, defined by \(b\). As a result, variations in \(a\) have a direct impact on the transverse electric field intensity, and consequently, on the efficiency of wakefield generation. In contrast, changes in \(b\) scale the transverse area without significantly modifying the field structure.

\vspace{1pc}

\noindent The pronounced increase in wakefield amplitude observed for the aspect ratio \(b/a = 0.70\) may reflect a geometric condition that enhances coupling between the microwave pulse and the plasma response. This behavior could be associated with a resonance-like effect or an optimal transverse confinement that promotes efficient excitation of the plasma wave. However, further analysis is necessary to assess the robustness of this feature under varying pulse parameters. These findings underscore the critical role of waveguide geometry in modulating the efficiency of wakefield excitation in microwave-driven plasma configurations.

\vspace{1pc}

\begin{figure*}
    \centering
    \includegraphics[width=0.85\linewidth]{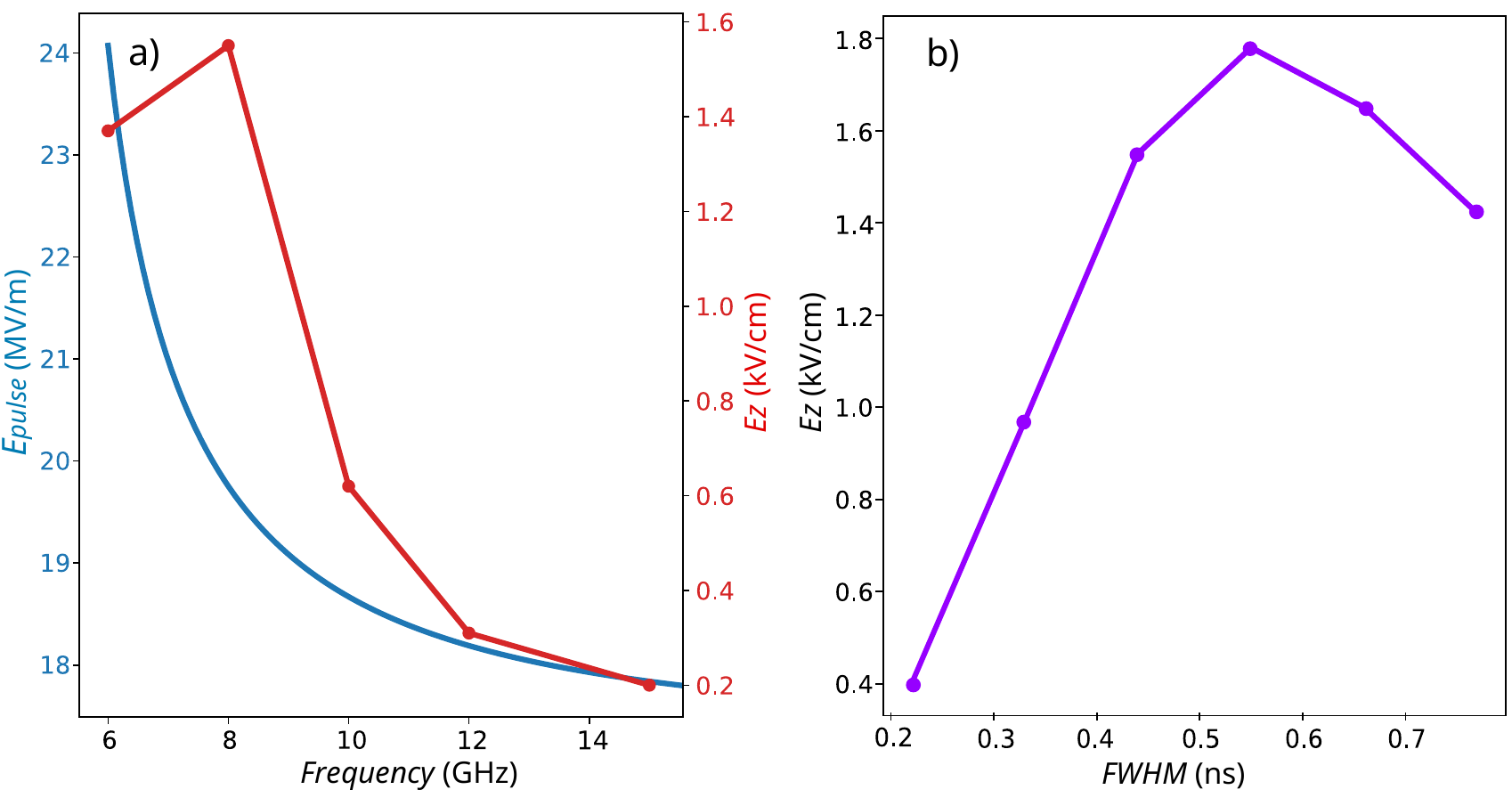}
    \caption{\label{fig:Ew_vs_dt}(a) Electric field amplitude of the microwave pulse, \(E_0\), (blue) and wakefield amplitude, \(E_z\), (red) as a function of the central pulse frequency \(f_0\). (b) Wakefield amplitude \(E_z\) as a function of the pulse duration \(\Delta t\) (FWHM). A maximum is observed around \(\Delta t \cong 0.55\)~ns, indicating resonant behavior in the excitation efficiency. Parameters: \(P = 0.25\,\text{GW}\), \(f_0 = 8\,\text{GHz}\), \(a = 3\,\text{cm}\), \(b/a = 0.7\).}
\end{figure*}

\noindent To evaluate the influence of the microwave pulse frequency on the efficiency of wakefield generation, a set of simulations was performed in which the central frequency \(f_0\) was varied from 6 to 15~GHz, while the pulse duration was maintained at approximately 0.44~ns. The waveguide geometry was held constant, with \(a = 3\,\text{cm}\) and \(b/a = 0.7\), and the peak power of the input pulse was fixed at 0.25~GW. As shown in FIG.~\ref{fig:Ew_vs_dt}(a), the wakefield amplitude \(E_z\) (red curve) exhibits a non-monotonic dependence on frequency: a maximum value of 1.55~kV/cm is reached at \(f_0 = 8\,\text{GHz}\), while both lower and higher frequencies result in significantly reduced wakefield amplitudes.

\vspace{1pc}

\noindent This behavior suggests the existence of an optimal frequency for coupling energy from the driving pulse into the electrostatic plasma response. Despite the increase in the number of carrier oscillations within the fixed pulse envelope at higher frequencies, the simulations reveal a reduction in wakefield amplitude. This behavior suggests that increasing the carrier frequency does not necessarily lead to more efficient wakefield generation. The observed trend may arise from phase mismatches or destructive interference between the driving pulse and the plasma response. In addition, the amplitude of the driving electric field is frequency-dependent for a given power. For the TE$_{10}$ mode in a rectangular waveguide, the field amplitude can be approximated as
\begin{equation}
    E_0 = \sqrt{ \frac{2P\, \omega \mu_0}{ab\, \beta} },
    \label{Eq_Eo}
\end{equation}
where \(P\) is the pulse power, \(\omega = 2\pi f_0\) is the angular frequency, \(\beta_{TE_{10}} = \sqrt{(\omega/c)^2 - (\omega_p/c)^2 - (\pi/a)^2}\) is the propagation constant, estimated using a dielectric plasma response characterized by the relative permittivity \(\epsilon_r = 1 - (\omega_p/\omega)^2\), and \(a\), \(b\) are the waveguide dimensions. This expression provides a reference estimate valid in the linear regime of cold, collisionless plasmas and serves to illustrate the frequency dependence of the field amplitude. As shown in FIG.~\ref{fig:Ew_vs_dt}(a), the field amplitude \(E_0\) (blue curve) slightly decreases with increasing frequency. This reduction results from the more rapid growth of \(\beta\) compared to \(\omega\), thereby lowering the driving field strength and diminishing the efficiency of wakefield excitation at higher frequencies.

\vspace{1pc}

\noindent The results indicate that strong plasma wakefield generation arises from a subtle interplay between pulse duration, carrier frequency, and plasma response. Wakefield excitation efficiency is not governed solely by the number of carrier cycles or the peak electric field amplitude. Instead, the existence of a favorable frequency regime is evident, with optimal excitation observed near \(f_0 = 8\,\text{GHz}\) under the present configuration. Beyond this point, the interaction becomes progressively less efficient.

\vspace{1pc}

\noindent It is also noteworthy that while the wakefield amplitude at \(f_0 = 6\,\text{GHz}\) remains comparatively high, it is still lower than at 8\,GHz. This behavior may be associated with the proximity of 6\,GHz to the cutoff frequency of the TE$_{10}$ mode, \(f_c \sim 5.0\,\text{GHz}\). Given the finite spectral bandwidth of the pulse, components near or below the cutoff are attenuated or fail to propagate, thereby reducing the effective energy delivered to the plasma. Moreover, as the frequency approaches cutoff, the group velocity decreases and dispersion increases, leading to temporal broadening of the pulse and a reduction in field gradient sharpness, both of which can diminish wakefield excitation.

\vspace{1pc}

\noindent The influence of pulse duration on wakefield excitation was also investigated. In this set of simulations, the full width at half maximum (FWHM) of the pulse envelope, denoted by \(\Delta t\), was systematically varied between 0.22 and 0.77~ns, while maintaining a constant central frequency \(f_0 = 8\,\text{GHz}\), fixed peak power of 0.25~GW, and unchanged waveguide dimensions. As shown in FIG.~\ref{fig:Ew_vs_dt}(b), the wakefield amplitude \(E_z\) exhibits a non-monotonic dependence on pulse duration, reaching a maximum of approximately 1.78~kV/cm at \(\Delta t = 0.55\,\text{ns}\), after which it gradually decreases with increasing duration. This behavior indicates a resonant-like response in the wakefield excitation efficiency as a function of \(\Delta t\). For short-duration pulses, the limited energy content and narrower temporal envelope result in a weaker ponderomotive force, resulting in weaker wakefields. Conversely, for longer pulses, the reduced field gradients and loss of temporal coherence diminish the energy coupling into the plasma, thereby lowering the amplitude of the excited wakefield.

\vspace{1pc}

\noindent The observed resonance around \(\Delta t \approx 0.55\,\text{ns}\) can be understood as the result of balancing several physical mechanisms. First, the temporal extent of the pulse must be sufficient to sustain a ponderomotive force over multiple plasma oscillations to enable coherent wave formation. Second, because the spectral width of the pulse scales inversely with duration, \(\Delta f \sim 1/\Delta t\), shorter pulses contain broader spectra, increasing their sensitivity to dispersion and cut-off effects. Third, increasing the pulse duration \(\Delta t\) results in a greater number of carrier oscillations within the envelope. Wakefield excitation is most effective when the number of cycles is sufficient to coherently drive the plasma response, without exceeding a threshold beyond which phase mismatches or destructive interference degrade the coupling efficiency. In the present configuration, optimal excitation occurs for approximately four to five cycles, corresponding to a central frequency of 8~GHz and a pulse duration of \(\Delta t = 0.55\,\text{ns}\).

\vspace{1pc}

\begin{figure*}[ht]
    \centering
    \includegraphics[width=0.90\linewidth]{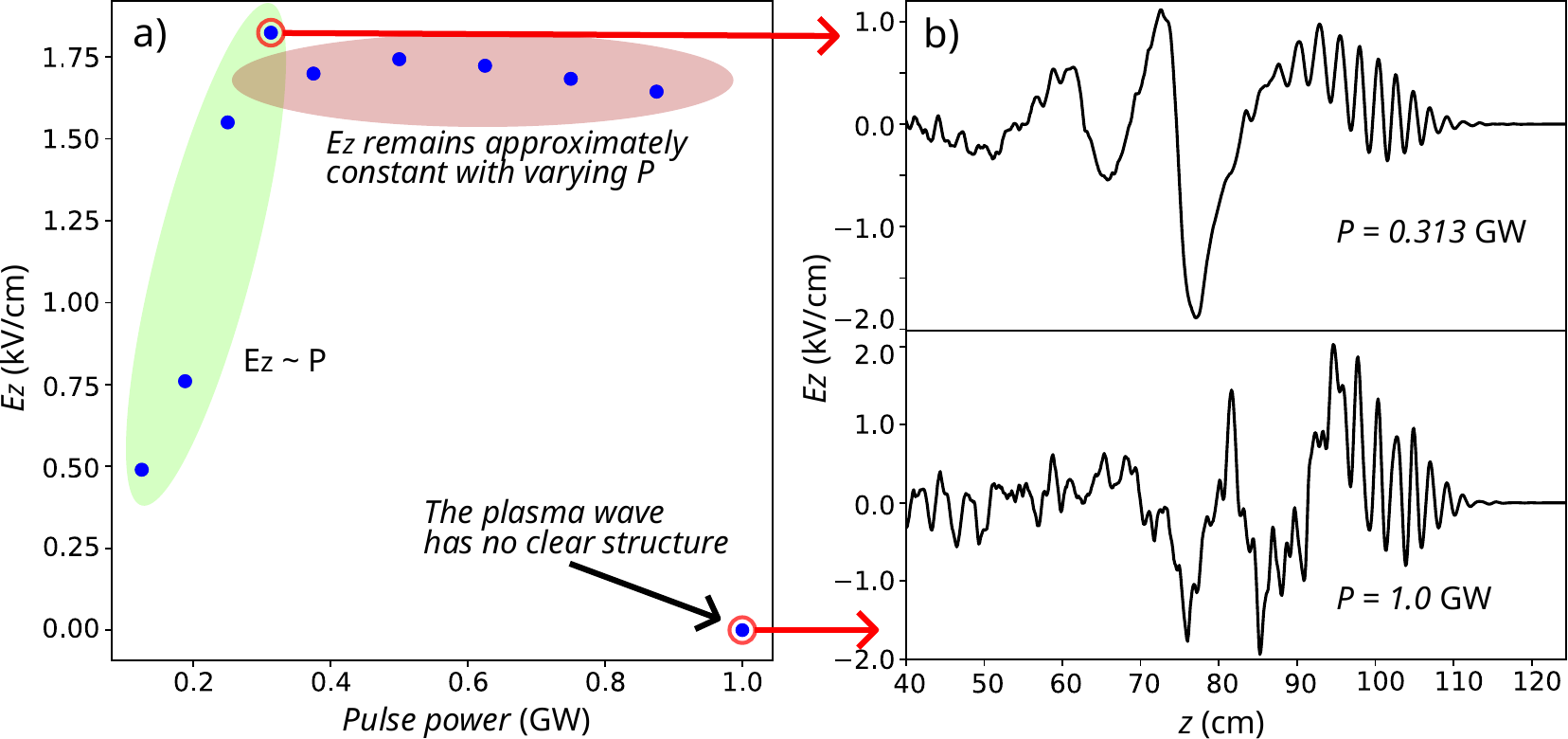}
    \caption{\label{fig:Ew_vs_Power} (a) Peak wakefield amplitude \(E_z\) as a function of the microwave pulse power \(P\). A rapid increase is observed at low powers, followed by saturation and eventual collapse at \(P = 1.0\)~GW. (b) Longitudinal electric field profiles \(E_z(z)\) along the waveguide axis (\(x = a/2\), \(y = b/2\)) for selected powers: 0.313 and 1.0~GW. The wake becomes strongest near 0.313~GW, saturates at 0.875~GW, and disappears at 1.0~GW, indicating the loss of coherent plasma response at high field strengths.}
\end{figure*}

\noindent To examine the impact of microwave pulse power on wakefield generation, simulations were performed in which the peak power \(P\) was varied from 0.125 to 1.0~GW, while all other parameters were held constant (\(f_0 = 8\,\text{GHz}\), \(\Delta t = 0.44\,\text{ns}\), \(a = 3\,\text{cm}\), \(b/a = 0.7\)). The results, presented in FIG.~\ref{fig:Ew_vs_Power}, show that the wakefield amplitude \(E_z\) increases rapidly with microwave power in the low-to-intermediate regime, reaching a maximum around \(P = 0.31\,\text{GW}\). Beyond this point, the amplitude saturates and fluctuates slightly, showing no further enhancement with increasing power. Remarkably, for pulse powers near 1.0~GW, the wakefield structure collapses entirely, with \(E_z\) effectively vanishing. At moderate power levels, increasing the pulse intensity leads to a larger electric field amplitude (see Eq.~(\ref{Eq_Eo})), thereby strengthening the ponderomotive force responsible for displacing plasma electrons and initiating wake formation. However, beyond a critical threshold, the interaction becomes highly nonlinear and sensitive to boundary effects. In particular, the excitation becomes impulsive: electrons are rapidly accelerated toward the waveguide walls and absorbed by the simulation’s boundary buffer before a coherent plasma wave can form. Consequently, the collective motion necessary to sustain the wakefield is suppressed. This effect is most pronounced at \(P \sim 1.0\,\text{GW}\), where the wake structure collapses.

\vspace{1pc}

\noindent These findings highlight the existence of an optimal power window for efficient wakefield excitation. While a minimum field strength is required to initiate the plasma response, excessively strong pulses hinder the development of the wake by causing particle losses, phase disruption, and field profile deformation. Properly balancing the pulse power is thus essential for maximizing wakefield strength and preserving the coherence of the excited plasma wave, an important consideration for the design of practical microwave-driven wakefield accelerators.

\vspace{1pc}

\noindent Finally, a set of simulations was carried out to assess the impact of plasma electron density \(n_e\) on wakefield generation. The density was varied by applying multiplicative factors to a reference value of \(n_0 = 1.8 \times 10^{16}\,\text{m}^{-3}\), while keeping all other parameters constant: \(P = 0.313\,\text{GW}\), \(f_0 = 8\,\text{GHz}\), \(\Delta t = 0.55\,\text{ns}\), \(a = 3\,\text{cm}\), and \(b/a = 0.7\). As reported in TABLE~\ref{tab:Ew_vs_density}, the wakefield amplitude \(E_z\) exhibits a monotonic increase with plasma density, rising from approximately 1.83~kV/cm at \(n_e = n_0\) to 2.46~kV/cm at \(n_e = 2n_0\). This trend is consistent with the dependence of the plasma frequency \(\omega_p\) on density, as the strength of the wakefield scales approximately as \(E_{wake} \sim \sqrt{n_e}\), see Eq.~(\ref{E_wake}). Higher density leads to a larger \(\omega_p\), leading more pronounced excitation of longitudinal electric fields. Although increasing the plasma density leads to enhanced wakefield amplitudes, this trend is fundamentally constrained by the requirement that the driving electromagnetic pulse must propagate through the plasma-filled waveguide. For the TE$_{10}$ mode in a rectangular geometry, the cutoff frequency can be estimated by setting the propagation constant \(\beta_{TE_{10}} = 0\), yielding
\( \omega_{\text{cutoff}} \approx \sqrt{\omega_p^2 + \left( \frac{\pi c}{a} \right)^2 } \). In the present simulations, all considered densities remain below this cutoff threshold, thereby enabling efficient pulse propagation and energy transfer to the plasma. Future investigations could explore regimes approaching this limit to delineate the boundaries of effective wakefield excitation in microwave-driven plasma systems.

\begin{table}
\caption{\label{tab:Ew_vs_density}Simulated wakefield amplitude \(E_z\) as a function of plasma electron density \(n_e\), varied as multiples of the base value \(n_0 = 1.8 \times 10^{16}\,\text{m}^{-3}\).}
\begin{ruledtabular}
\begin{tabular}{cccc}
 & \(n_e/n_o\) & \(E_z\) (kV/cm) & \\
\hline
 & 1.0 & 1.83 &  \\
 & 1.5 & 2.05 &  \\
 & 2.0 & 2.46 &  \\
 & 2.5 & 2.65 &  \\
 & 3.0 & 3.35 &  \\
 & 3.5 & 3.38 &  \\
\end{tabular}
\end{ruledtabular}
\end{table}
\section{Summary and Conclusions}
\noindent In this work, we performed a comprehensive three-dimensional Particle-in-Cell (PIC) study of plasma wakefield generation driven by microwave pulses propagating through plasma-filled rectangular waveguides. The simulations systematically explored the influence of key physical parameters, including waveguide geometry, pulse frequency, duration, peak power, and plasma density, on the amplitude and structure of the resulting electrostatic wakefield.

\noindent The present study demonstrates, through fully electromagnetic three-dimensional PIC simulations, that longitudinal electrostatic fields, i.e., plasma wakefields, can be excited in rectangular waveguides filled with low-density plasma using TE$_{10}$ microwave pulses. While previous works have shown the feasibility of wakefield generation in similar configurations, our results provide a comprehensive analysis of how key physical parameters, including waveguide geometry, pulse duration, frequency, power, and plasma density, modulate the amplitude and spatial structure of the wake. This expanded understanding not only supports the viability of microwave-driven wakefield schemes, but also offers quantitative guidance for optimizing such systems in future experimental implementations.

\noindent The study revealed that the transverse dimensions of the waveguide, especially its width \(a\), play a dominant role in determining the wakefield amplitude. Wider guides lead to decreased energy density and weaker wakes, whereas variations in the aspect ratio \(b/a\) had minimal impact, except for a resonant-like enhancement observed at \(b/a = 0.7\) when using a central frequency pulse of 8~GHz. These findings are consistent with the spatial field profile of the TE$_{10}$ mode.

\noindent In terms of pulse parameters, a resonant-like dependence of the wakefield amplitude on both frequency and duration was observed. Under the specific simulation conditions employed in this study, the optimum frequency for wakefield excitation was found to be \(f_0 = 8\,\text{GHz}\), beyond which the amplitude decreased due to reduced electric field strength and increased phase mismatch. Similarly, an optimal pulse duration of approximately 0.55\,ns was identified, beyond which the wakefield amplitude declined as a result of diminished field gradients and reduced temporal coherence.

\noindent The peak power of the driving pulse influences the wakefield response in a nonlinear manner. In the parameter range explored, an optimal power level near 0.31\,GW was identified, beyond which the wakefield amplitude saturated or slightly decreased due to nonlinear plasma effects.

\noindent Finally, increasing the plasma density resulted in stronger wakefields, in agreement with theoretical scaling laws based on the plasma frequency. However, this favorable trend is ultimately limited by the waveguide cut-off condition and plasma opacity at high densities.

\noindent Altogether, this work provides a detailed numerical foundation for the design of microwave-driven wakefield accelerators, identifying optimal parameter regimes and guiding future experimental implementations in plasma-based compact acceleration systems.
\section*{Acknowledgments}
\noindent The authors thank the Universidad Industrial de Santander (UIS), Colombia, for supporting this work through internal funding (project ID:~3706).
\section{REFERENCES}
\nocite{*}
\bibliography{aipsamp}
\end{document}